# New polymorph γ-CuWO$_4$ inspired by γ-CuMoO$_4$: experimental identification and theoretical verification

Jiri Houska[*], Stanislav Haviar, Jiri Capek, Radomir Cerstvy, Kalyani Shaji, Nirmal Kumar, Petr Zeman

*Department of Physics and NTIS - European Centre of Excellence,
University of West Bohemia, Univerzitni 8, 301 00 Plzen, Czech Republic*
[*]*corresponding author, email jhouska@kfy.zcu.cz*

**Abstract**

In the context of our efforts to develop hydrogen gas sensors, two samples of ternary CuWO$_4$ with the same crystalline structure have been prepared by two different non-equilibrium techniques. We show that the materials' structure is significantly different from the known stable one, and we identify that it is very similar to that of previously reported γ-CuMoO$_4$. We use ab-initio calculations to confirm that the newly identified phase, γ-CuWO$_4$, represents a local energy minimum. We present very similar calculated and measured lattice constants and X-ray diffractograms. We make a case that the low-temperature formation of the metastable γ-CuWO$_4$ phase is facilitated by easier kinetics and/or by Cu-rich composition of our samples, and we show that it converts to the stable phase after annealing to 600 °C.

**Keywords**

CuWO$_4$; WO$_3$; CuO; nanoparticles; sensors

**Main text**

There are worldwide efforts [1,2] to continuously improve the performance of semiconductor metal oxide-based gas sensors, for various applications in the field of hydrogen economy, environmental monitoring, breath monitoring or food analysis. The sensors are of particular interest at the nanometer scale, where the large surface-to-volume ratio significantly improves their sensitivity. Further progress has been achieved by moving from homogeneous metal oxides to nanocomposites, nanostructures, heterojunctions and core-shell nanoparticles (NPs), which combine two semiconductors such as n-type WO$_3$ and p-type CuO [3-12]. Sensing of a wide range of gases has been reported using this pair of materials, ranging from hydrogen [12] and hydrogen sulfide [10] through ammonia [6] or acetone [5] to xylene [3] or sarin [9].

In this context, we are making efforts to prepare WO$_3$+CuO based sensors by two different deposition techniques which realize two different designs: (i) film of WO$_3$ on nanoparticles of CuO and (ii) alternating monolayers of nanoparticles of WO$_3$ and nanoparticles of CuO. In both cases, the research includes annealing of the as-deposited sensors to various temperatures in synthetic air. While the sensing performance of the materials and devices obtained will be published separately, the present paper deals with one of these materials itself. We have found that at a temperature of at most 400 °C, a new polymorph, referred to as γ-CuWO$_4$ throughout this paper, formed at the interface of WO$_3$ and CuO. The structure of this polymorph is significantly different from that of the known stable phase which is usually given without any Greek letter and which we refer to as α-CuWO$_4$ for distinguishing purposes. However, the structure of the newly identified polymorph is very similar to the known γ-CuMoO$_4$ phase.

The aim of this paper is to examine the existence of the newly identified polymorph γ-CuWO$_4$, to specify under which conditions did it form, to confirm that it constitutes a local energy minimum according to ab-initio calculations, and to check whether it converts to stable α-CuWO$_4$ after annealing to a sufficiently high temperature. In addition to a possible usefulness of γ-CuWO$_4$ itself, the presented information is expected to be useful for the correct interpretation of results obtained by the numerous practitioners dealing with the aforementioned devices based on WO$_3$+CuO.

The sample based on a combination of a thin film of $WO_3$ on CuO NPs was prepared in two steps. First, the NPs were prepared utilizing a magnetron-based gas aggregation source (MGAS). The MGAS (HVM Plasma) was attached to a main deposition chamber, where a silicon substrate was mounted to a rotating substrate holder kept at a room temperature. The MGAS featured of a cylindrical magnetron equipped with a 50 mm diameter Cu target placed inside a cylindrical aggregation chamber measuring 100 mm in diameter, terminated by a conical ending with a 4 mm diameter orifice. The NPs were prepared in Ar atmosphere at an Ar flow rate of 100 sccm and a pressure of 80 Pa. The magnetron was driven in DC mode and operated at a constant discharge power of 100 W. Second, 40 nm thin film of $WO_3$ was deposited on top of the NPs using another magnetron-based sputtering system (Leybold-Heraeus LH Z400). This film was deposited using a magnetron with a 72 mm diameter W target powered by a DC power supply operated at a constant power mode of 60 W. The total pressure of the Ar + $O_2$ working atmosphere was 614 mPa and the corresponding flow rates of Ar and $O_2$ were 15.0 and 3.75 sccm, respectively. During the deposition, the sample was heated to 400 °C. Let us note that the NPs were prepared as metal NPs during the first step, but they oxidezed to CuO during their exposure to the ambient atmosphere when the sample was transferd to the second system and also during the deposition process of the $WO_3$ film. More details about the experimental setups can be found in Refs. [12,13].

The sample based on alternating monolayers of $WO_3$ and CuO NPs was deposited using another MGAS (Nanogen-Trio, Mantis deposition) equipped with three 1" magnetrons placed inside a cylindrical aggregation chamber with a diameter of 120 mm, terminated by a domed ending with a 4 mm diameter orifice. In this experiment, Cu and W metallic targets were used. This MGAS was also attached to a main deposition chamber, where a silicon substrate was mounted to a rotating substrate holder kept at a room temperature. The NPs were prepared in Ar + $O_2$ atmosphere at the total pressure of 60 Pa. The Ar flow rate was 100 sccm. The flow rates of $O_2$ were 1.3 and 1.5 sccm for $WO_3$ and CuO NPs, respectively. Both magnetrons were driven with DC power supply (GEN 600-1.3, TDK-Lambda) operated at a constant discharge power of 25 W. The deposition rate was regularly monitored during the deposition by crystal microbalance (STM-2, Inficon). This enabled switching among the deposition of copper oxide and tungsten oxide NPs in a precisely defined ratio. The system was orchestrated with a custom-built LabVIEW system. More details about this system can be found in Ref. [14].

Both kinds of samples were post-annealed in a furnace in synthetic air at an ambient pressure at 400, 500 and 600°C for 6 hours.

The structural characterization was performed by X-ray diffraction measurements, using an X'Pert PRO (PANalytical) instrument with Cu K$\alpha$ radiation ($\lambda$ =0.154187 nm). All data evaluation was performed using the PANalytical ware package HighScore Plus. The evaluation included the Rietveld analysis: least-square fit of lattice vectors of $\gamma$-$CuWO_4$ on its measured X-ray diffractogram, using calculated crystal coordinates of atoms.

All ab-initio density-functional theory simulations were performed by the Quantum Espresso code [15]. The main simulation algorithm was a simultaneous relaxation of all lattice vectors and all 36 atomic positions of $\gamma$-$CuWO_4$, using those of $\gamma$-$CuMoO_4$ as a starting point. Two simulation techniques were employed. First, the exchange and correlation energy was represented by the PBE (Perdew–Burke–Ernzerhof) functional [16], the atom cores and inner electron shells were represented by standard Vanderbilt-type ultrasoft pseudopotentials, the energy cutoff was 30 Ry and the electronic density cutoff was 360 Ry. The optimization was controlled by convergence threshold on total energy ($10^{-4}$ Ry), convergence threshold of forces ($10^{-3}$ Ry/bohr) and, most importantly, convergence threshold on |stress| (10 MPa, compared to ≈10 GPa after the replacement of Mo by W in $\gamma$-$CuMoO_4$; the last criterion to be fulfilled). Second, the exchange and correlation energy was for comparative purposes represented also by the PBEsol (PBE revised for solids) functional [17], the corresponding Vanderbilt (W) and PAW (projector augmented wave; Cu and O) pseudopotentials were taken from the online database [18], the energy cutoff was 90 Ry and the electronic density cutoff was 720 Ry. Reducing the total stress to less than 200 MPa (approaching or being below usual stresses in experimentally prepared films) turned out to be sufficient to conclude that the difference between measured and calculated lattice parameters is for both simulation techniques smaller than the difference between these two calculated results.

All simulations were spin-polarized. The lowest-energy ferrimagnetic configuration of γ-CuWO$_4$ (see below) has been identified before and used as a part of the starting point of the variable-cell geometry optimization, by performing the initial fixed-cell geometry optimization 6× with 6 different starting points. The starting points included a non-magnetic configuration and 5 ferrimagnetic configurations. Two of these six initial configurations (one inspired by γ CuMoO$_4$, and another one which converged to the same magnetic state after some of the atoms changed the sign of their magnetic moments) led to the identical final configuration with the lowest energy. Similar simulations of the known α-CuWO$_4$ phase in its lowest-energy antiferromagnetic configuration were performed for comparison.

The X-ray diffraction patterns of both samples prepared are shown in Figure 1. As indicated above, the interdiffusion of binary amorphous WO$_3$ and CuO after annealing to 400 °C leads to a formation and crystallization of ternary CuWO$_4$, characterized by diffraction patterns very far from that of conventionally expected triclinic α-CuWO$_4$ (PDF #04-009-6293; space group $P\bar{1}$). The difference cannot be explained by parallel crystallization of other phases such as CuO (PDF #00-048-1548; space group C2/c). However, the diffraction patterns obtained are very similar to those of triclinic γ-CuMoO$_4$. Note that there are two γ-CuMoO$_4$ PDF records #01-088-0619 and #04-009-2227 (space group $P\bar{1}$), very similar except different choices which vector is labelled a/b/c in the corresponding primary source (we use the labelling from the latter one). Furthermore, the atomic coordinates in γ-CuMoO$_4$ according to powder diffraction files [19] and according to the widely used online database [20] can be converted to each other after replacing some coordinates by their complement to 1 and after exchanging Cu and Mo atoms in the x-direction.

Thus, a case can be made that γ-CuMoO$_4$ is a prototype of a newly created metastable γ-CuWO$_4$ phase. The similarity of the 36-atom primitive cells and in turn X-ray diffractograms is supported by the fact that 30 atoms are the same and the remaining 6 atoms are not only in the same group, but within ≈0.5% also of the same size. Compare the lattice constants of bcc-Mo of 3.147 Å and bcc-W of 3.165 Å, and note that the extra shell of valence electrons of W is almost compensated by 14 extra protons (the period of W includes lanthanoids, the period of Mo does not) attracting these electrons to the atomic core.

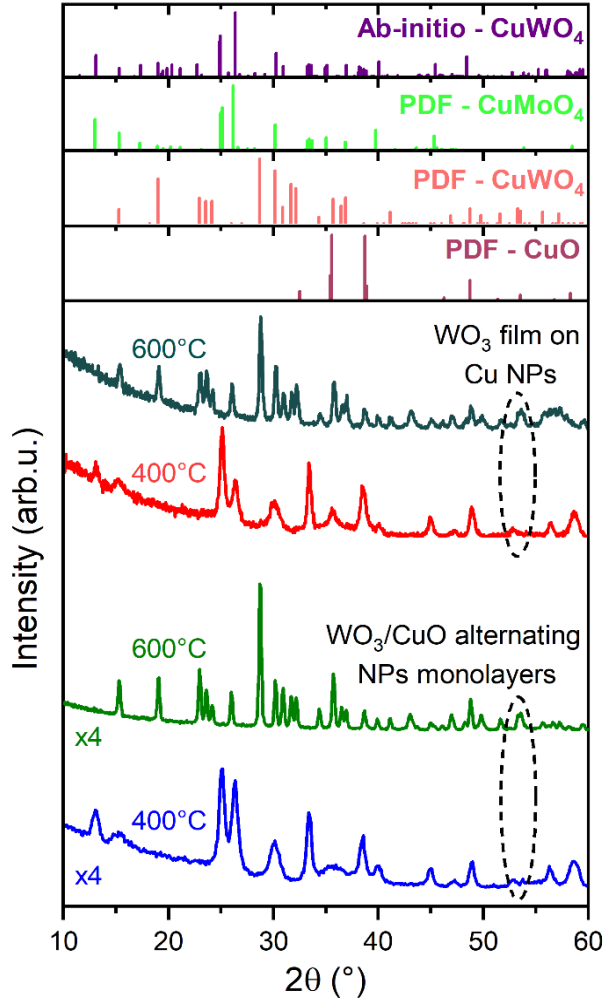

**Figure 1**. The first and third line from the bottom show X-ray diffractograms (Cu Kα radiation) of samples annealed to 400 °C, dominated by the newly identified metastable γ-CuWO$_4$ phase. See the calculated peak positions and relative intensities in the top line, similar to those of γ-CuMoO$_4$ (PDF #01-088-0619 or equivalently #04-009-2227; space group $P\bar{1}$). The second and fourth line from the bottom show X-ray diffractograms of samples annealed to 600 °C, dominated by the known stable α-CuWO$_4$ phase (PDF #04-009-6293; also space group $P\bar{1}$). There are two kinds of samples: WO$_3$ film on CuO NPs (top dashed oval) and alternating monolayers of WO$_3$ NPs and CuO NPs (bottom dashed oval). Although the whole samples are prone to be Cu-rich, the differences in diffractograms cannot be explained by a presence of CuO (PDF #00-048-1548; space group C2/c).

**Table 1**. Characteristics of triclinic (space group $P\bar{1}$ in all cases) CuMoO$_4$ and CuWO$_4$ phases mentioned in the text.

| Material | Source | $a$ (Å) | $b$ (Å) | $c$ (Å) | $\alpha$ (°) | $\beta$ (°) | $\gamma$ (°) | Atoms per cell | Volume per atom (Å$^3$) |
|---|---|---|---|---|---|---|---|---|---|
| α-CuWO$_4$ | Experiment (PDF #04-009-6293) | 4.703 | 5.839 | 4.878 | 91.68 | 92.47 | 82.81 | 12 | 11.061 |
| γ-CuMoO$_4$ | Experiment (PDF #04-009-2227) | 6.306 | 7.978 | 9.710 | 103.27 | 103.21 | 94.76 | 36 | 12.726 |
| γ-CuWO$_4$ | Experiment (Rietveld analysis) | 6.351 | 8.029 | 9.688 | 102.77 | 104.08 | 94.81 | 36 | 12.843 |
| γ-CuWO$_4$ | Ab-initio (PBE functional) | 6.356 | 8.121 | 9.888 | 103.57 | 103.27 | 94.60 | 36 | 13.278 |
| γ-CuWO$_4$ | Ab-initio (PBEsol functional) | 6.288 | 7.935 | 9.711 | 102.71 | 103.24 | 94.56 | 36 | 12.660 |

In order to confirm that γ-CuWO$_4$ constitutes at least a local energy minimum, ab-initio simulations have been performed. The simulations were complemented by the Rietveld analysis of one of the X-ray diffractograms obtained. All resulting lengths of lattice vectors and angles between them are provided in Table 1. It can be seen that all four sets of available values (γ-CuWO$_4$ according to the Rietveld analysis, γ-CuWO$_4$ according to the simulations using PBE and PBEsol functional and even γ-CuMoO$_4$) are very similar. The values of $a$, $b$ and $c$ are within ±1.2%, the values of $\alpha$, $\beta$ and $\gamma$ are within ±0.4% and the values of cos$\alpha$, cos$\beta$ and cos$\gamma$ are within ±3.2%. Furthermore, the differences of $a$, $b$ and $c$ between given lines are predominantly in the same direction, preserving the shape of the primitive cell. For example, the values of $a$, $b$ and $c$ of γ-CuWO$_4$ obtained using the PBE functional (implementation of the generalized gradient approximation, known to overestimate rather underestimate lattice constants) are 100-102% of those obtained by the Rietveld analysis. The values of $a$, $b$ and $c$ obtained using the PBEsol functional (specifically tailored to improve the structural properties of solids) are 99-100% of those obtained by the Rietveld analysis. Thus, the presented measured and calculated data (i) mutually support each other and (ii) support the statement that the prototype of the γ-CuWO$_4$ material obtained is γ-CuMoO$_4$. Note that the differences between modelling and experiment are actually smaller than the mutual differences between both modelling techniques used. For example, the calculated volume per atom is 103.4% (PBE functional) and 98.6% (PBEsol functional) of the experimental one, at a difference between PBE and PBEsol of 4.8%.

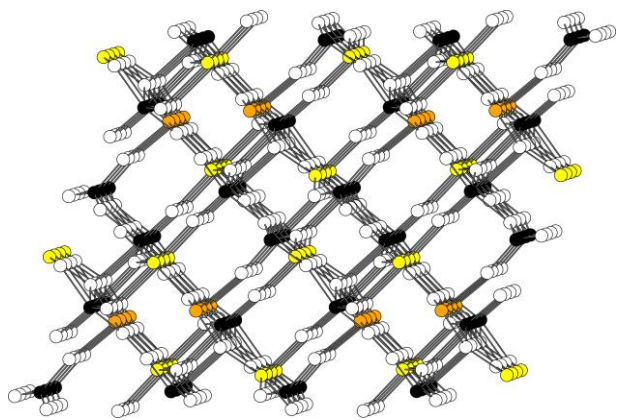

**Figure 2**. Calculated (top line in Fig. 1) structure of triclinic (space group $P\bar{1}$) γ-CuWO$_4$. Yellow and orange balls are Cu atoms carrying magnetic moments of opposite signs, black balls are W atoms and white balls are O atoms. There are 4×2×2 36-atom primitive cells and the view direction is slightly off x-axis.

The atomic structure of γ-CuWO$_4$ is visualized in Fig. 2, and the corresponding atomic coordinates are given in Table 2. See the trilayer structure consisting of almost planar $(0, \bar{1}, 1)$ layers (from the top left to the bottom right corner of Fig. 2) of CuWO$_3$, CuWO$_3$ and O$_2$. Moreover, there are triplets of almost planar (0,1,0) layers of Cu atoms carrying magnetic moments of opposite signs (horizontal in Fig. 2; 1× orange Cu+ and 2× yellow Cu-) and being responsible for the ferrimagnetism of γ-CuWO$_4$. Note the different local environment of Cu+ and Cu- atoms in terms of O-Cu-O angles. Next, the calculated lattice vectors and atomic coordinates were used to predict an X-ray diffractogram using an arbitrary full width at half maximum of all peaks of 0.1414°, and this diffractogram is compared with one of the experimental ones in Fig. 3. It is clear that despite the numerous sources of potential differences (finite exactness of ab-initio calculations, off-stoichiometric composition in the experiment, preferred orientation in the experiment), the diffractograms are very similar.

**Table 2**. Calculated (top line in Fig. 1) crystal coordinates of atoms in the γ-CuWO$_4$ phase. There are 36 atoms in a primitive cell of a space group $P\bar{1}$, i.e. each line represents two atoms with crystal coordinates $x,y,z$ and $-x,-y,-z$. The signs of Cu+ and Cu- represent ferrimagnetism.

| Atom | $x$ | $y$ | $z$ |
|---|---|---|---|
| Cu+ | 0.073395 | 0.470897 | 0.666430 |
| Cu- | 0.221178 | 0.233297 | 0.436402 |
| Cu- | 0.596684 | 0.199555 | 0.003486 |
| W | 0.090052 | 0.116342 | 0.892554 |
| W | 0.431846 | 0.455677 | 0.227344 |
| W | 0.720302 | 0.127631 | 0.348393 |
| O | 0.001728 | 0.077441 | 0.273005 |
| O | 0.046785 | 0.283652 | 0.806048 |
| O | 0.059110 | 0.700784 | 0.521798 |
| O | 0.139408 | 0.896054 | 0.022150 |
| O | 0.191711 | 0.456619 | 0.296525 |
| O | 0.291206 | 0.024408 | 0.548759 |
| O | 0.320761 | 0.244237 | 0.061347 |
| O | 0.343198 | 0.414444 | 0.614311 |
| O | 0.372489 | 0.031136 | 0.845849 |
| O | 0.397693 | 0.603530 | 0.118059 |
| O | 0.470569 | 0.227883 | 0.355280 |
| O | 0.711132 | 0.339653 | 0.200659 |

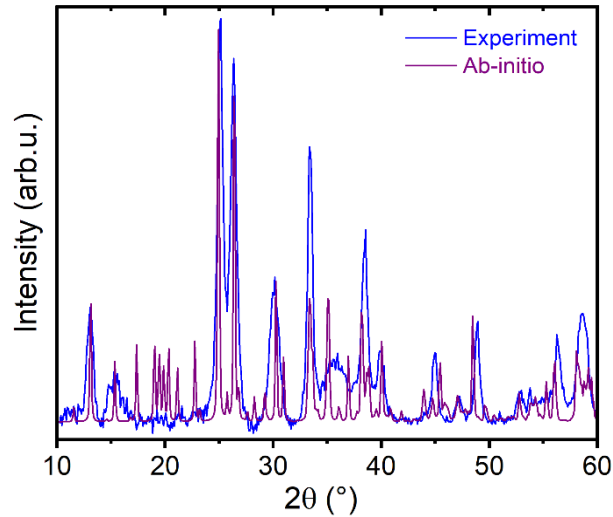

**Figure 3**. Comparison of an experimental X-ray diffractogram (Cu Kα radiation; blue; bottom line from Fig. 1 after subtracting the background) and a predicted X-ray diffractogram (purple; top line from Fig. 1 after converting the delta functions to peaks with a full width at half maximum of 0.1414°).

Let us recall the metastable nature of γ-CuWO$_4$. This has been also confirmed by ab-initio calculations (PBE functional): the formation energy of α-CuWO$_4$ (-1.563 eV/at. compared to fcc-Cu + bcc-W + O$_2$) is of 5 meV/at. lower than that of γ-CuWO$_4$ (-1.558 eV/at.). Thus, it is relevant to discuss possible reasons for the formation of this metastable phase. First, let us point out the overall non-equilibrium nature of the preparation techniques such as sputtering. The importance of that is further supported by the fact that the aforementioned energy difference of 5 meV/at is actually rather low, compared to energy differences between polymorphs of many other materials.

Second, as the samples analyzed were not prepared in order to contain α- or γ-CuWO$_4$ but in order to be good sensors, their overall [Cu]/[W] compositional ratio is higher than 1. In particular, about the same volume of alternating nanoparticles of WO$_3$ (≈54 Å$^3$ per metal atom) and nanoparticles of CuO (≈21 Å$^3$ per metal atom) leads to a higher number of Cu atoms. Although the agreement of modelling and experiment (Table 1 and Fig. 3) indicates that the [Cu]/[W] ratio in the γ-CuWO$_4$ phase (which constitutes only part of the samples) is very close to 1, the preference to form γ-CuWO$_4$ rather than α-CuWO$_4$ may be nevertheless enhanced under Cu-rich conditions. We made efforts to support also this hypothesis by ab-initio calculations. Indeed, when one of the six W atoms in the 36-atom cell of γ-CuWO$_4$ and in a 1×1×3 36-atom supercell of α-CuWO$_4$ is replaced by Cu, the energy of γ-CuWO$_4$, originally of 5 meV/at. higher, becomes of ≈5 meV/at. lower than that of α-CuWO$_4$. The key information is not the latter number itself (one arbitrary composition, possible role of distribution of Cu atoms in the W sublattice, poorer convergence for the fictitious Cu-rich composition than for the ideal one), but the direction in which the energy difference changes.

Third, note that γ-CuWO$_4$ is of ≈16% less dense phase than α-CuWO$_4$ (volumes per atom in Table 1). Thus, it is reasonable to assume that while α-CuWO$_4$ is slightly more thermodynamically stable, the low-temperature preparation of γ-CuWO$_4$ is significantly kinetically easier. This is a well known phenomenon reported for other materials (to put one example: easier growth of low-density metastable γ-Al$_2$O$_3$ compared to high-density stable α-Al$_2$O$_3$). The low-temperature preparation may be quantified by the following: metastable γ-CuWO$_4$ formed during the annealing up to 400 °C as shown by two of the diffractograms in Fig. 1, remained stable up to at least 500 °C (not shown), but converted to stable α-CuWO$_4$ after annealing to 600 °C (possibly in parallel to changing its composition from slightly Cu-rich to stoichiometric - previous paragraph) as shown by the other two diffractograms in Fig. 1.

To conclude, a new polymorph γ-CuWO$_4$, having a triclinic lattice (space group $P\bar{1}$) very similar to its γ-CuMoO$_4$ prototype, has been obtained by annealing the boundary between WO$_3$ and CuO. Ab-initio simulations confirm the existence of this phase, confirm its slight metastability compared to α-CuWO$_4$, and allowed us to predict the atomic positions and in turn an X-ray diffractogram very similar to the experimental one. While the formation of γ-CuWO$_4$ during the annealing to 400 °C could be supported by its lower density and possibly also by slightly Cu-rich composition, it converts to α-CuWO$_4$ when annealed to 600 °C. In addition to a possible usefulness of γ-CuWO$_4$ itself, the findings are expected to contribute to a correct interpretation of results obtained in the course of worldwide efforts to prepare gas sensors of various architectures combining WO$_3$ and CuO.

This work was supported by the project Quantum materials for applications in sustainable technologies (QM4ST), funded as project No. CZ.02.01.01/00/22_008/0004572 by Programme Johannes Amos Comenius, call Excellent Research. Computational resources were provided by the e-INFRA CZ project (ID:90254), supported by the Ministry of Education, Youth and Sports of the Czech Republic.